\documentclass[twocolumn,nopacs,amsmath,amssymb,prl,superscriptaddress,floatfix]{revtex4}

\usepackage{graphicx}
\usepackage{epsfig}
\usepackage{bm}

\begin{document}

\title{Defect-induced supersolidity with soft-core bosons}
\author{F. Cinti}
\email{cinti@sun.ac.za}
\affiliation{Max Planck Institute for the Physics of Complex Systems, N\"othnitzer Str. 38, 01187 Dresden, Germany}
\affiliation{National Institute for Theoretical Physics (NITheP), Stellenbosch 7600, South Africa}
\author{T. Macr\`i}
\affiliation{Max Planck Institute for the Physics of Complex Systems, N\"othnitzer Str. 38, 01187 Dresden, Germany}
\author{W. Lechner}
\affiliation{IQOQI and Institute for Theoretical Physics, University of Innsbruck, Austria}
\author{G. Pupillo}
\affiliation{IPCMS (UMR 7504) and ISIS (UMR 7006), Universit\'e de Strasbourg and CNRS, Strasbourg, France}
\author{T. Pohl}
\affiliation{Max Planck Institute for the Physics of Complex Systems, N\"othnitzer Str. 38, 01187 Dresden, Germany}

\begin{abstract}
More than 40 years ago, Andreev, Lifshitz, and Chester suggested the possible existence of a 
peculiar solid phase of matter, the microscopic constituents of which can flow superfluidly 
without resistance due to the formation of zero-point defects in the ground state of 
self-assembled crystals. Yet, a physical system where this mechanism is unambiguously 
established remains to be found, both experimentally and theoretically. Here we investigate 
the zero-temperature phase diagram of two-dimensional bosons with finite-range soft-core 
interactions. For low particle densities, the system is show to feature a solid phase in which 
zero-point vacancies emerge spontaneously and give rice to superfluid flow of particles 
through the crystal. This provides the first example of defects-induced, continuous-space 
supersolidity consistent with the Andreev-Lifshitz-Chester scenario. 
\end{abstract}


\maketitle
Spontaneous symmetry breaking is a focal principle of condensed matter physics -- yet,
simultaneous breaking of fundamentally different symmetries represents a rare 
phenomenon. A prime example is the so-called supersolid phase \cite{bop12}, which displays both crystalline and superfluid properties, that is, the simultaneous breaking of continuous translational and global gauge symmetry. The first mentioning of such a state goes back to 
Gross \cite{gro57}, who predicted the possibility of a density-modulated superfluid phase of weakly interacting bosons described by a classical field. 
Later, Andreev, Lifshitz \cite{anl69}, and Chester (ALC) \cite{che70} conjectured a microscopic mechanism for strongly interacting systems based on two key
assumptions: first, that the ground state of a bosonic crystal contains defects such as vacancies and interstitials, and, second, that these defects 
can delocalize, thereby, giving rise to superfluidity. 
However, the physical realizability of this scenario has since remained under active debate \cite{bal10}.\\
\indent In 2004, torsional oscillator experiments \cite{kic04a,kic04b} provided first suggestive evidence for superfluidity in solid $^4$He through a rapid drop of the resonant oscillation period below a critical temperature, viewed indicative for superfluid decoupling of a fraction of the He crystal. This finding has sparked a host of new experimental activity \cite{suso_exp0,suso_exp1,suso_exp2,suso_exp3,suso_exp4,suso_exp5,suso_exp9,suso_exp10,suso_exp6,suso_exp7,suso_exp8}, that, however, challenged the original interpretation and pointed out several artefacts causing a non-supersolid origin of the observations. 
Theoretical work has established that crystal incommensurability is a necessary condition for superfluidity 
\cite{prs05} and that zero-point defects in ground state solid He are prevented by a large activation energy 
\cite{ceb04,bkp06}. In addition, Boninsegni \textit{et al.} \cite{bkp06} Rota and Boronat \cite{rob12}, Ma \textit{et al.} \cite{npt08}, and Lechner and Dellago \cite{led09} have shown that point-like defects 
experience an effective attraction that results in defect-clustering and phase separation, ruling out the possibility
of defect-induced supersolidity \cite{bkp06} as in the ALC scenario. 
Several experiments \cite{suso_exp3,suso_exp9,suso_exp10,suso_exp6} have shown that the original observations were caused by shear 
modulus stiffening of bulk solid He, and later found no signature of superfluidity on avoiding this effect \cite{kic12}. As a result, there now seems to be consistent experimental and theoretical evidence for the absence of the long-sought supersolid phase in He. The mere existence of continuous-space supersolidity induced by zero-point defects thus remains an open question.

\begin{figure}[t!]
\begin{center}
\resizebox{0.99\columnwidth}{!}{\includegraphics{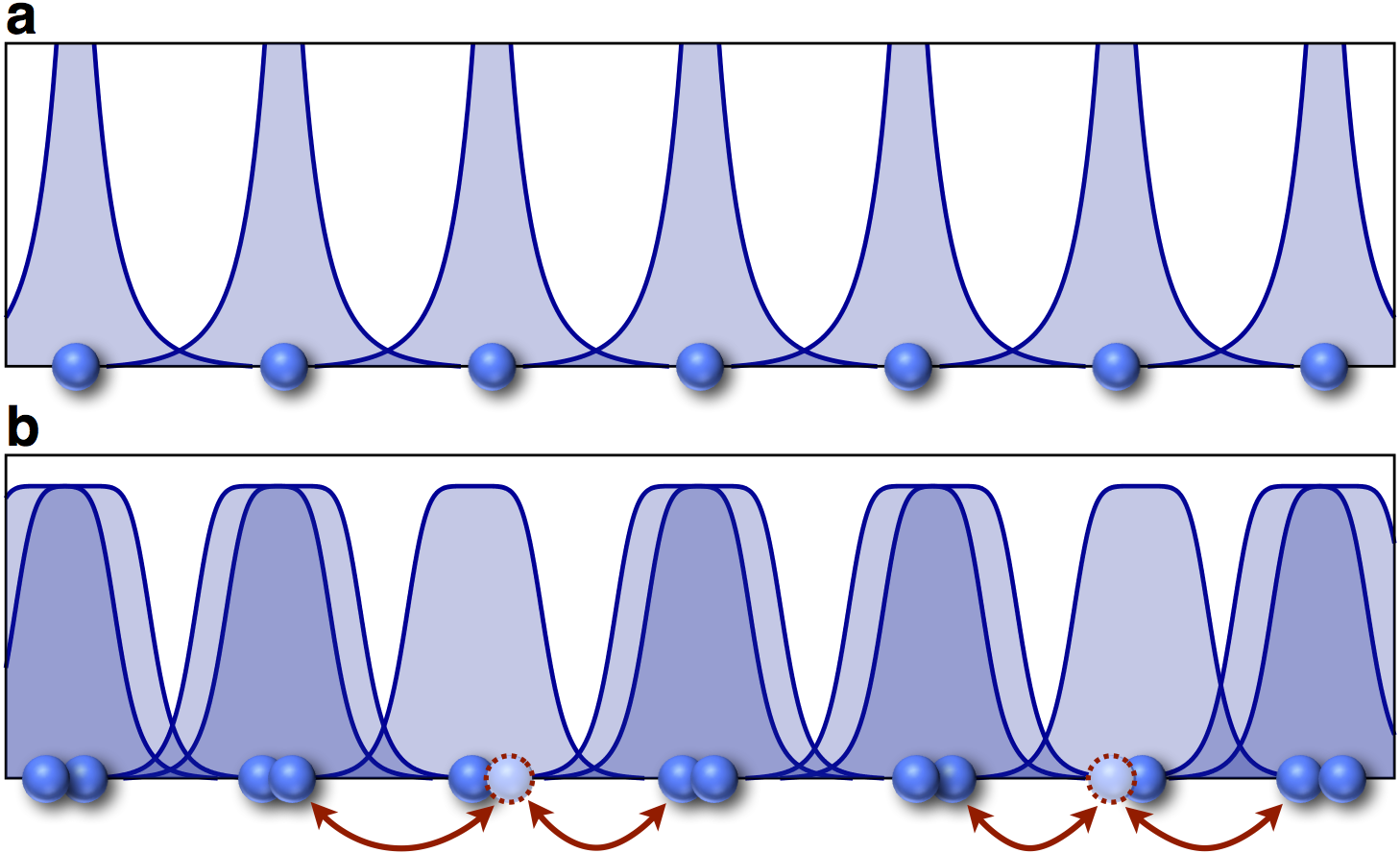}}
\caption{\label{fig1}{\bf Crystallization due to singular and soft-core interactions.} (a) For singular potentials with pure power-law repulsion, indicated by the blue areas, particles assemble into a commensurate and insulating solid. (b) Upon removing the singularity, particles can cluster and, under proper conditions, form an incommensurate crystal with more particles than lattice sites. Defect delocalization due to inter-site tunneling can then promote a finite superfluid response of the self-assembled crystalline ground state.}
\end{center}
\end{figure}

In this Article, we show that bosonic particles interacting via soft-core potentials (see Fig.\ref{fig1}) provide a prototype system for addressing this question. Using exact numerical techniques, we determine the underlying zero-temperature phase diagram, which reveals the emergence of defect-induced supersolidity in the vicinity of commensurate solid phases, as conjectured by ALC  \cite{anl69,che70}.\\

\section{Results}
{\bf Supersolidity with Soft-core bosons.}
We consider a two-dimensional ensemble of $N$ bosons with density $\rho$, interacting via a pair potential of the type 
\begin{equation}\label{eq:pot}
V=\frac{V_0}{r^\gamma+R_{\rm c}^\gamma}.
\end{equation}
This interaction approaches a constant value $V_0/R_{\rm c}^\gamma$ as the inter-particle distance, $r$, decreases below the soft-core distance $R_{\rm c}$, and drops to zero for $r>R_{\rm c}$. The limiting case $\gamma\rightarrow\infty$ yields the soft-disc model \cite{por94}, while $\gamma=3$ and $\gamma=6$ correspond to soft-core dipole-dipole \cite{cjb10} and van der Waals \cite{hnp10} interactions that can be realized with ultracold atoms \cite{hnp10,mhs11} or polar molecules \cite{bdl07,mpb07}. Here, we focus on the latter case ($\gamma=6$), for which the Hamiltonian reads
\begin{equation}\label{eq:ham}
\hat{H}=-\sum_{i=1}^{N}\frac{\nabla^2_i}{2}+\sum_{i<j}^N\frac{U}{1+r_{ij}^6},
\end{equation}
where the units of length and energy are $R_{\rm c}$ and $\hbar^2/mR_{\rm c}^2$, respectively, and $m$ denotes the particle mass. In these units, the zero-temperature physics is controlled by the dimensionless interaction strength $U=mV_0/(\hbar^2R_{\rm c}^4)$ and the dimensionless density $R_{\rm c}^2\rho$.\\
\indent Particles with soft-core interactions have been studied previously in the field of soft condensed matter physics \cite{lik01,mgk06,csk11}, in the classical high-temperature regime. One of the main findings has been that pair potentials with a negative Fourier component \cite{lik01} favor the formation of particle clusters, which in turn can crystallize to form a so-called cluster-crystal. In the quantum domain, theoretical work has so far focused on the regime of weak interactions and high particle densities \cite{hnp10,cjb10,sjr10,smb11,kuk12,mjr12}, which was shown to be well described by mean-field calculations \cite{hcj12,mmc12}. In this limit, one finds strongly modulated superfluid states \cite{gro57,por94,hnp10,cjb10} with broken translational symmetry in the form of a density-wave. 

\begin{figure}[t!]
\begin{center}
\resizebox{0.99\columnwidth}{!}{\includegraphics{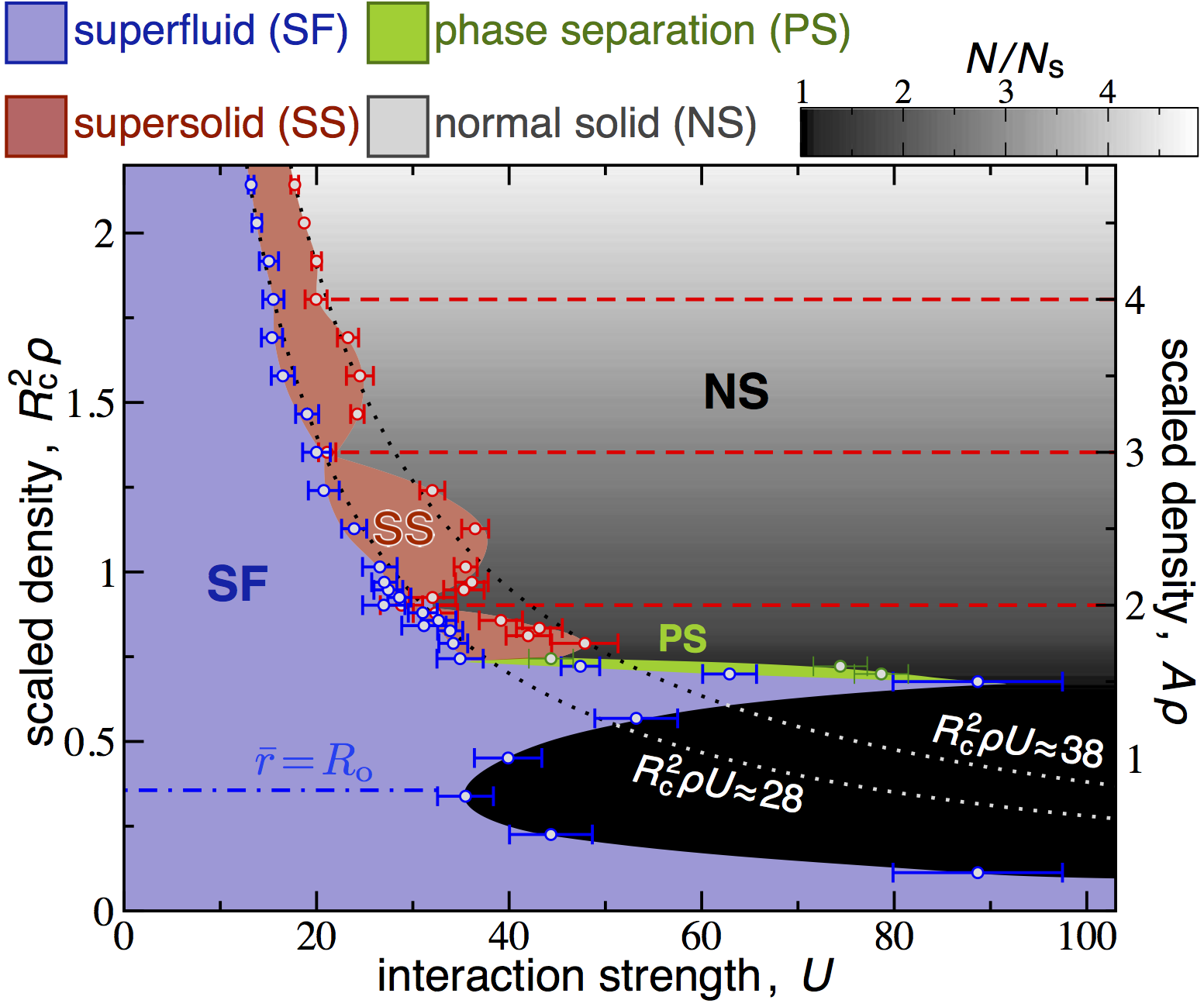}}
\caption{\label{fig2}{\bf Zero-temperature phase diagram of two-dimensional soft-core bosons.} 
The phase diagram displays the emergence of superfluid (SF) and different solid (NS) and supersolid (SS) 
phases for varying interaction strength $U$ and density $\rho$. The density on the left $y$-axis has been 
scaled by the soft-core radius $R_{\rm c}$. The right axis gives the density in units of the inverse area, 
$A=\sqrt{3}(1.6R_{\rm c})^2/2$, of the unit cell of the high-density solid phase, corresponding to the lattice site 
occupation $N/N_{\rm s}$ for a given number of particles and lattice sites, $N$ and $N_{\rm s}$, respectively. 
For $A\rho\gtrsim 1.5$, the grey region labeled as NS corresponds to a cluster crystal with $N/N_{\rm s}>1$, 
as indicated by the grey scale. Supersolid phases with 
different occupation numbers are found between two hyperbolas, defined by $R_c^2 \rho U={\rm const.}$ (dotted lines). 
At high densities ($A\rho\gtrsim3.5$) they can be understood in terms of density-modulated superfluids. In contrast, 
superfluidity within the low-density supersolid lobes emerges from delocalized zero-point defects according to the ALC scenario.
The horizontal error bars represent statistical uncertainties and uncertainties due to the finite stepping of $U$.}
\end{center}
\end{figure}

In the following we investigate strong coupling domain where correlations and quantum fluctuations are expected to become important. We employ path integral Monte Carlo simulations to determine the ground state properties of the Hamiltonian eq.(\ref{eq:ham}) (see Methods section). 
\begin{figure*}[t!]
\begin{center}
\resizebox{0.99\textwidth}{!}{\includegraphics{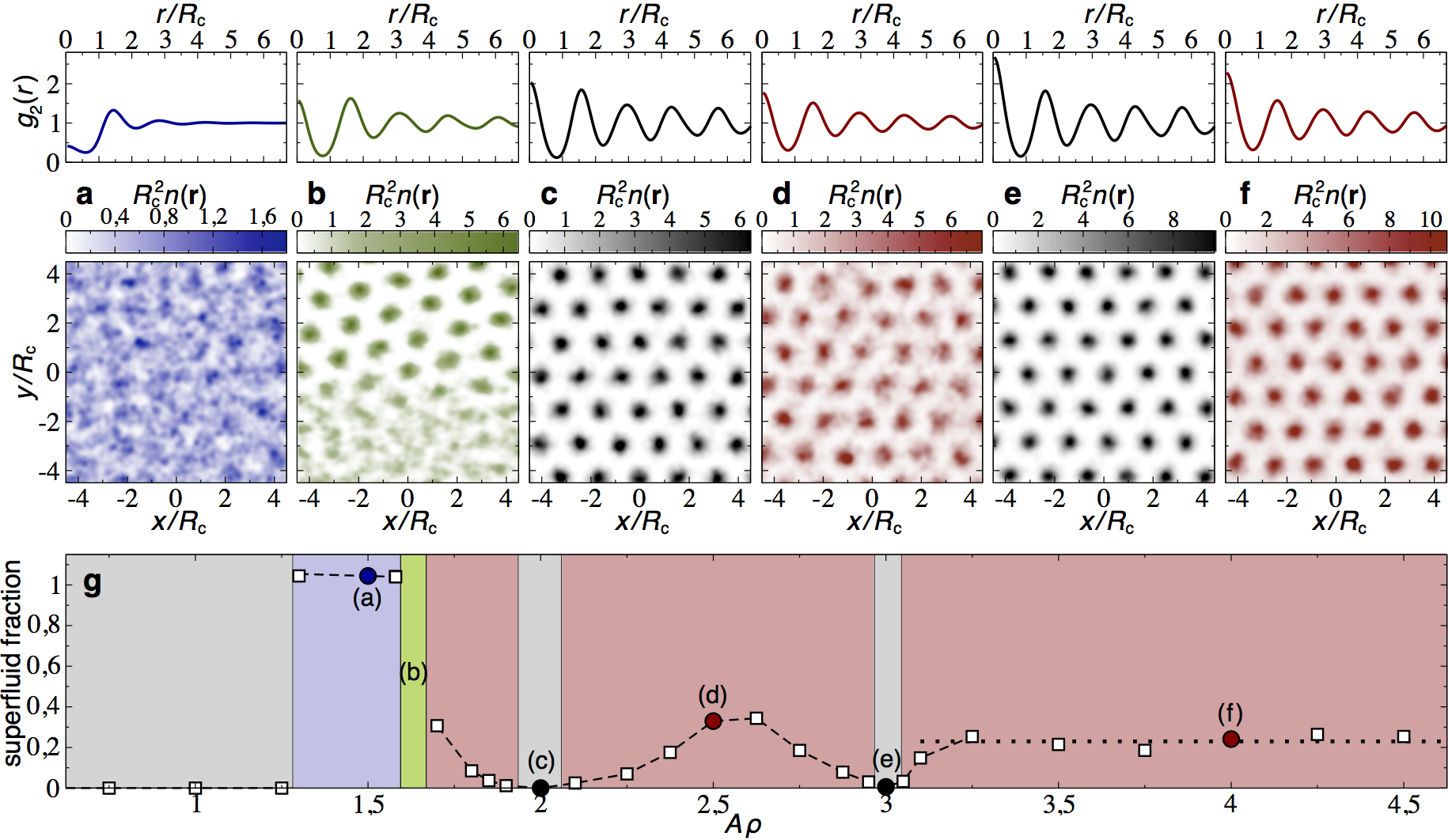}}
\caption{\label{fig3} {\bf Characterization of the different phases across the phase diagram.} Panels (a)-(f) show the radial pair correlation function $g_2(r)$ and the particle density distribution $n({\bf r})$ for $\rho R_0^2 U=32$ and different values of $A\rho$, indicated in (g). Panel (g) displays the superfluid fraction $f_s$ as a function of $A\rho$, covering all phases indicated in Fig.\ref{fig2}. Global superfluidity is not well defined in the phase-separation region (green area) and therefore omitted. The horizontal dotted line shows the $\rho\rightarrow\infty$ limit of $f_{\rm s}$ for a density-wave supersolid, obtained from a
mean-field evaluation following the approach of Leggett \cite{leg70}.}
\end{center}
\end{figure*}
The obtained phase diagram, shown in Fig.\ref{fig2}, reveals a rich spectrum of phases with varying interaction 
strength and density.\\ 

\noindent{\bf Small particle densities.} At  small densities $R_{\rm c}^2\rho\lesssim 0.5$ we find two phases: a superfluid and an 
insulating triangular crystal composed of singly occupied sites, that is, where the number of lattice sites, $N_{\rm s}$, equals the particle number $N$.
The observed lobe structure of this crystalline region is readily understood by noticing that at very low densities, that is, large inter-particle distances 
$\bar{r}=1/\sqrt{\pi \rho}>R_{\rm c}$, the physics is dominated by the long-range tail of the interaction potential, 
$V\sim1/r^6$. For a fixed interaction strength $U\gtrsim 35$, we thus find a first order  liquid-solid 
quantum phase transition with increasing $A\rho$, consistent with previous work on bosons with power-law 
interactions \cite{bdl07,abk07,mpw07,oal11}. In particular, the location of the liquid-solid phase transition for very low densities coincides with that for pure van der Waals interactions. With increasing density, however, the average inter-particle spacing, $\bar{r}$, approaches the soft-core radius $R_{\rm c}$ and drops to values for which equation (\ref{eq:pot}) strongly deviates from pure $\sim1/r^6$ interactions and levels off below the turning point $R_{\rm o}=(5/7)^{1/6}R_{\rm c}$ . As a result of the decreasing repulsive inter-particle forces, the crystal melts again for increasing densities. 
As indicated in Fig.\ref{fig2}, we indeed find a re-entrant superfluid at particle densities for which $\bar{r} < R_{\rm o}$.\\

{\bf Intermediate densities.} A distinctive consequence of the soft-core interaction is that the energy cost for forming close particle pairs is 
bound by $V_0$. This potentially enables the formation of crystalline phases with $N > N_{\rm s}$ above a critical 
density where doubly occupied lattice sites become energetically favorable on increasing the lattice constant. 
As expected for a triangular crystal, the lattice constant decreases as $a=(\sqrt{3}\rho/2)^{-1/2}$ at small 
densities. However, around $a\approx1.4\,R_{\rm c}$ it increases again and settles to a 
density-independent value of $a_{0}\simeq1.6\,R_{\rm c}$ upon further increase of $\rho$. 
The corresponding volume of the unit cell 
$A=\sqrt{3/4}a_0^2$ provides a measure of the lattice occupancy $N/N_{\rm s}=A\rho$, which is also shown in 
Fig.\ref{fig2}. The transition to cluster crystals occurs at $A\rho\approx1.5$. This indeed coincides with the 
critical density for crystallization of the reentrant superfluid phase.\\
\indent Around this density, a thin region of phase separation is found to lay in between the cluster crystal and the superfluid phase. Fig.\ref{fig3}(b) shows a typical example for the particle density distribution in this region. Two distinct coexisting phases can be recognized: a crystal phase with exactly two particles per site (upper part of the figure), and a superfluid phase (lower part). We have carefully checked that the occurrence of this phase-separated state is not an artifact of the simulations, by performing accurate annealing and by choosing different initial conditions, such as random and different crystalline configurations.\\
\indent Above incommensurate lattice occupations $N/N_{\rm s}\gtrsim1.5$, the direct liquid-solid quantum phase 
transition is replaced by a first order transition from a superfluid to a supersolid phase. The supersolid phase is 
approximately found to occur between the two hyperbola defined by $R_{\rm c}^2\rho U=\alpha$, 
with $\alpha\approx28$ and $\alpha\approx38$, respectively (see dotted lines in Fig.\ref{fig2}). 
These two lines are derived from the weak 
interaction limit ($U\rightarrow0$ and $\rho\rightarrow\infty$ with $\alpha={\rm const.}$), where mean-field theory 
predicts a transition to a density-wave supersolid that is determined only by the value of $\alpha$ (Ref. \cite{hnp10}). 
While this mean-field prediction becomes exact in the high-density limit (see Fig.\ref{fig2}), the situation 
is dramatically different at moderate densities where the discrete nature of the particles plays a significant role. 
This gives rise to the emergence of supersolid regions with a lobe structure, which vanish at commensurate lattice 
occupations $N/N_{\rm s}=2$ and $N/N_{\rm s}=3$. There, we find a direct transition between a superfluid and an 
insulating solid phase.\\

\begin{figure}[t!]
\begin{center}
\resizebox{0.99\columnwidth}{!}{\includegraphics{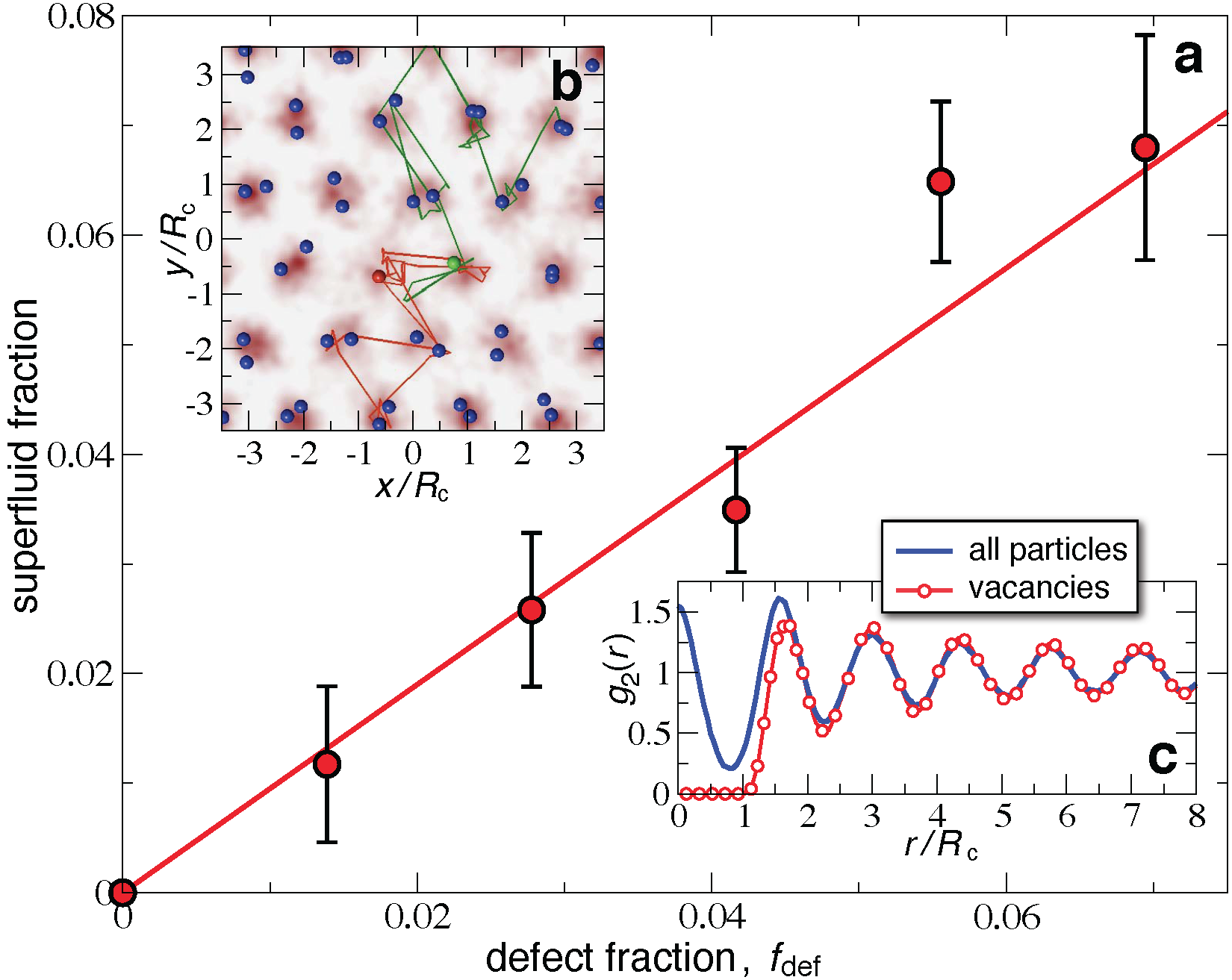}}
\caption{\label{fig4} {\bf Defect-induced superfluidity.} Panel (a) shows the superfluid fraction in a doubly occupied solid ($N/N_{\rm s}=2$) 
as a function of the fraction of defects in the form of singly occupied sites for fixed $U=31$ and varying particle number between $N=139$ and $N=144$. For $N=144$ the system forms an insulating commensurate solid with $N_{\rm s}=72$ doubly occupied sites. Decreasing the particle number does not affect $N_{\rm s}$ but leads to the formation of a small fraction $f_{\rm def}=(2N_{\rm s}-N)/N_{\rm s}$ of singly occupied lattice sites.
Panel (b) shows the particle density (colour code) for $f_{\rm def}=0.03$, obtained from a Monte Carlo configuration along with the particle positions (spheres) for a single imaginary-time slice. The two initially adjacent vacancies (red and green sphere) delocalize, as indicated by the corresponding imaginary-time trajectories. Accordingly, the vacancy-vacancy pair correlation function (red), shown in (c) for $f_{\rm def}=0.07$, resembles the $g_2(r)$ of the underlying crystal (blue), as expected for a delocalized, very dilute gas of repulsive bosons.  
The error bars in panel a represent statistical uncertainties of the Monte Carlo sampling.}
\end{center}
\end{figure}

\indent This behavior is illustrated in Fig.\ref{fig3}, where the superfluid fraction, $f_{\rm s}$, is shown as a function of 
$A\rho$ for $\alpha=32$, that is, in between the two dotted lines in Fig.\ref{fig2}. As shown in Fig.\ref{fig3}(c) 
and (e), at $A\rho=2$ and $A\rho=3$ one finds a commensurate crystal with exactly $N/N_{\rm s}=2$ 
[Fig.\ref{fig3}(c)] and $N/N_{\rm s}=3$ [Fig.\ref{fig3}(e)] particles per lattice site, respectively, and vanishing superfluidity 
[Fig.\ref{fig3}(g)]. Importantly, the crystal structure in between these two densities is practically unchanged, as 
seen by comparing the particle density distributions $n({\bf r})$ and density-density correlation functions $g_{2}(r)$ 
in Figs.\ref{fig3}(c)-(e). However, the incommensurate lattice filling $N/N_{\rm s}$ and the resulting fluctuations of 
individual site occupations enables particles to tunnel between the sites. This gives rise to a non-vanishing 
superfluid fraction of the crystal, which can assume sizable values of 
$f_{\rm s}=0.3$ for $N/N_{\rm s}\approx A\rho=2.5$.\\

{\bf High densities.} For higher densities $A\rho>3$, the scenario described above changes considerably. As shown in Fig.\ref{fig3}(g) 
the superfluid fraction approaches a constant, density-independent value $f_{\rm s}\approx0.24$ with increasing 
$\rho$. In particular, for an average commensurate filling $N/N_{\rm s}=4$ there is no direct phase transition 
between a superfluid and a solid insulating phase, and instead the supersolid phase persists with no significant 
difference to the case of incommensurate lattice occupancies $N/N_{\rm s}\neq4$. This behavior signals a 
crossover to the regime where the supersolid phase can be understood in terms of a density modulated superfluid 
\cite{gro57,por94,hnp10,cjb10,hcj12}, where the discrete nature of the particles becomes irrelevant. In this 
limit, the superfluid-supersolid quantum phase transition is well captured by a mean-field description 
\cite{por94,hnp10,mmc12}. It predicts a transition point at $\alpha=R_{\rm c}^2\rho U=28.2$ as well as a 
superfluid fraction that is solely determined by the value $\alpha$, and yields $f_{\rm s}=0.23$ for $\alpha=32$. As shown in Figs.\ref{fig2} and 
\ref{fig3}(g) both predictions are well confirmed by our Monte Carlo results for $A\rho\gtrsim3.5$, suggesting that the transition to 
density-wave supersolidity takes place at a surprisingly small number of only $N/N_{\rm s}\approx3.5$ particles per lattice site.\\

{\bf Defect delocalization.} The most interesting behavior takes place around the superfluid-solid quantum phase transition at 
$N/N_{\rm s}=2$. Figure \ref{fig4} provides a more detailed 
look at the transition between the insulating crystal and the supersolid phase, that is, for $U=31$ and $N/N_{\rm s}\approx2$. Starting from the insulating solid 
with doubly occupied lattice sites, we successively remove a small number of particles from randomly chosen 
sites and monitor the superfluid fraction of the resulting new ground state obtained from our simulations.
Removing a small number of particles does not cause structural changes of the ground state but rather creates 
a small fraction $f_{\rm def}=(2N_{\rm s}-N)/N_{\rm s}$ of zero-point crystal defects in the form of singly occupied
sites. An analysis of the Monte Carlo configurations shows that defects do not cluster and instead delocalize, as 
illustrated in Fig.\ref{fig4}b. This is also confirmed by the vacancy-vacancy pair correlation function, shown in Fig.\ref{fig4}c. For $r\gtrsim a_0$ it  closely resembles the $g_2(r)$ of the underlying solid, as expected for a very dilute gas of repulsive bosons. Indeed, we find a finite superfluid fraction even for small defect 
concentrations, which increases linearly with $f_{\rm def}$. We have verified that this finding is pertinent to 
the ground state and not to a metastable configuration by performing simulations with different initial conditions, including clustered defects. The observed behavior is, thus, consistent with defect-induced supersolidity according to the ALC scenario, and constitutes the central result of this work.

\section{Discussion}
Supersolidity in this system is the consequence of two unique features of soft-core bosons. First, the energy cost for 
forming close particle pairs is bound by $V_0$, which facilitates the formation of cluster crystals that naturally 
entail zero-point defects. Second, the dynamics and interaction of these defects differs fundamentally from those of 
conventional solids. In the latter case, vacancies and interstitials induce displacement fields that lead to purely 
attractive defect interactions \cite{bkp06,npt08,led09,led09b} and, therefore, prevent a delocalization of defects 
\cite{bkp06}. In cluster solids, on the other hand, defect interactions are purely repulsive, since they 
interact via the same underlying particle interaction $V(r)$ of eq.~\eqref{eq:pot}. In the present case, the 
transition between these two regimes is controlled by the particle density. For $A\rho\gtrsim1.5$ delocalized 
zero-point defects allow for the formation of supersolid phases. Below this density particles do not explore the 
soft core part of the interaction potential, such that defects are attractive and supersolidity is absent 
consistent with the results of Boninsegni \textit{et. al.} \cite{bkp06}. Around the transition region $A\rho\approx 1.5$ 
neither picture applies, and one observes separation between a superfluid and a doubly occupied, 
insulating cluster solid. Preliminary calculations 
based on path integral Langevin dynamics \cite{langevin} suggest that in this region structural and dynamical 
heterogeneity can give rise to a quantum glass phase at finite temperature.\\
\indent Having identified a physical system that facilitates defect-induced supersolidity, we hope that this work will provide useful guidance for future experiments and initiate further theoretical explorations. An important question concerns the general features of the interaction potential that are required to maintain the type of supersolid states described in this work. While the emergence of density-wave supersolids is largely insensitive to the detailed shape of the soft-core interaction, the low-density physics described in this work may be strongly affected. In fact, it seems reasonable to expect an interesting competition between intra- and inter-site interactions within the self-assembled crystal that will strongly depend on the long-range tail of the particle interactions. Moreover, the role of the dimensionality and confined geometries of finite systems represent another outstanding issue, and in particular their role for frustration effects with regards to defect delocalization.\\ 
\indent While the considered interactions do not straightforwardly occur in natural crystals, they can be designed in ultracold atom experiments. Recent experiments with Bose-Einstein condensates in optical cavities have already demonstrated a density-wave supersolid due to the breaking of a discrete translational symmetry \cite{bgb10,mbb12} and theoretical work has devised several schemes \cite{bdl07,mpb07} for manipulation of long-range interactions between polar molecules by external fields. Moreover, far off-resonant excitation of high lying Rydberg states \cite{hnp10,pmb10,mhs11} in degenerate atomic gases \cite{hrb08,vbr11,sce12} was shown to realize interactions of the type of equation (\ref{eq:pot}). Following Maucher \textit{et. al} \cite{mhs11}, such a Rydberg dressing of $^{87}$Rb condensates to Rb($35p_{3/2}$) states \cite{tfs04} with a laser detuning of $\sim500$MHz and an intensity of $~100$kW/cm$^2$ would produce a sizeable interaction strength of $U\sim35$. While we have focussed here on zero-temperatures, we have also performed finite-temperature simulations, showing that these parameters will permit the experimental observation of defect-induced supersolid phases for temperatures $T\lesssim10$nK, around typical densities $\sim10^8$cm$^{-2}$ and with a condensate lifetime of $\sim30$ms, limited by radiative decay of the weakly admixed Rydberg state. Recent experimental breakthroughs reporting the first observation of Rydberg interaction effects in a laser-driven Bose-Einstein condensate \cite{bkg13} hold high promise for the near-future realization of the setting  described in this work.
\section{Methods}
{\small{{\bf Numerical Details.}
Our numerical results were obtained from path-integral Monte Carlo simulations \cite{cep95} based on the continuous-space worm algorithm \cite{bon06} to determine the equilibrium properties of equation (\ref{eq:ham}) in the canonical ensemble, that is, at a fixed temperature temperature $T$ and a fixed particle number, chosen between $N=100$ and $N=400$. From these simulations we obtain, for example, density profiles, $n({\bf r})=\langle\sum_i\delta({\bf r}-{\bf r}_i(t))\rangle_t$, and pair correlation functions, $g_2(r) = [2 \pi n (N-1) r]^{-1} \langle\sum_{i}\sum_{j\neq i} \delta (r-r_{ij}(t)) \rangle_t$ as well as the superfluid fraction $f_s$, computed from the area estimator, as described in Sindzingre \textit{et al.} \cite{skc89} and Pollock \textit{et al.} \cite{poll87}. Here, $r_{ij}=|{\bf r}_j-{\bf r}_i|$, ${\bf r}_i$ are the positions of the $i=1,...,N$ particles, and $\langle..\rangle_t$ denotes an average of the corresponding imaginary time trajectories ${\bf r}_i(t)$. The properties of the system ground state were obtained by extrapolating to the limit of zero temperature, that is, by lowering the temperature until observables, such as the total energy, superfluid fraction and pair-correlations did not change upon further decrease of $T$. Moreover, the size of our two-dimensional simulation box with periodic boundary conditions was varied to assure insensitivity to system size. The calculated observables were used to construct the phase diagram. The first order superfluid-normal solid transition is detected by an abrupt increase of the maximum value ($S_{\rm max}$) of the static structure factor $S(k)=1+\rho\int{\rm d}{\bf r}{\rm e}^{i{\bf k}{\bf r}}(g_2(r)-1)$ and a simultaneous vanishing of the superfluid fraction. The superfluid-supersolid transition is characterized by a jump of $S_{\rm max}$ and an abrupt decrease of the superfluid fraction from $f_{\rm s}\approx1$ to a finite value $f_{\rm s}>0$, while the supersolid-normal solid transition is signaled by the vanishing of $f_{\rm s}$.
}}

\section{Acknowledgement}
We thank M. Boninsegni, S. Pilati, N. V. Prokof'ev, and S. G. S\"oyler for valuable discussions. 
This work was supported by the EU through the ITN COHERENCE.  G. P. is supported by the ERC-St Grant ``ColDSIM'' 
(grant agreement 307688) and EOARD. W.L. acknowledges support by the Austrian Science Fund through P 25454-N27.

\end{document}